\documentclass[aps,twocolumn,prd,showpacs,nofootinbib]{revtex4}
\usepackage{amsmath}
\usepackage{graphicx}
\usepackage{dcolumn}
\usepackage{bm}
\usepackage{amssymb}
\usepackage{latexsym}

\def\be{\begin{equation}}
\def\ee{\end{equation}}
\def\ba{\begin{eqnarray}}
\def\ea{\end{eqnarray}}

\begin{document}

\title{Emergently Thermalized Islands in the Landscape}

\author{Yun-Song Piao}

\affiliation{College of Physical Sciences, Graduate School of
Chinese Academy of Sciences, Beijing 100049, China}


\begin{abstract}

In this note, we point out that in the eternal inflation driven by
the metastable vacua of the landscape, it might be possible that
some large and local quantum fluctuations with the null energy
condition violation can stride over the barriers between different
vacua and straightly create some islands with radiation and matter
in new vacua.
Then these thermalized islands will evolve with the standard
cosmology.
We show that
such islands may be consistent with our observable universe, while
has some distinctly observable signals, which may be tested in
coming observations.




\end{abstract}


\maketitle

Recently, the string landscape with large number of metastable
vacua has received increased attentions \cite{BP, KKLT, S}, which
generally exhibit the cosmological dynamics as eternal inflation
\cite{V1983, L1986}. The tunnelling between various vacua of
landscape may be mediated by the CDL instanton \cite{CDL}, which
behaves as a bubble nucleating in new vacuum. However, the bubble
universe nucleated is generally empty and negatively curved, which
can hardly become our real world. Thus to produce an universe
containing the structure, a period of slow roll inflation
inside the bubble is needed \cite{FKMS}, which
reduces the curvature, provides the primordial density
perturbation and large number of entropy required by the
observable universe. The occurrence of inflation with enough
efolding number requires that the potential above new minimum
should have a nearly flat and long plain, which actually means a
fine tuning, since the regions with nearly flat potential are
generally expected to be quite rare in the string landscape. Thus
it will be interesting to check whether there are other
possibilities to lead to an universe like ours.

The landscape will be populated during the eternal inflation. Thus
in principle there may be many quantum fluctuations with various
spatial and temporal scales in each vacua of the landscape, which
may violate the NEC \cite{W}, see also Ref. \cite{W0612} for
discussions. However, most of these fluctuations are cosmological
irrelevant, since generally they will be generated and then return
rapidly to their original vacuum. The significant fluctuations are
those enough large, as studied in the island universe model
\cite{DV, Piao0506}, in which the vacuum background is that with
the observed value of the current cosmological constant. Thus with
diverse environments of string landscape, we may expect that it is
possible that some large and local quantum fluctuations with the
NEC violation can stride over the barrier between different vacua
in the landscape and straightly create some thermalized regions in
new vacua. This might be phenomenally illustrated by applying the
HM instanton \cite{HM}, since the HM instanton may be regarded as
a thermal fluctuation whose rate can be given by the difference in
the entropies between the fluctuation and the equilibrium state
\cite{KKLT}. In a normal landscape in which the barriers between
various minima are neither sharp nor broad \cite{BK}, the CDL
instanton and HM instanton will be expected to coexist
\cite{L0611}. This suggests that in some regions the bubbles with
new vacua will be nucleated, while in other regions the jump to
the top of the potential barrier will occur, which is mediated by
the HM instanton, and then the field will rapidly roll down along
the another side of barrier to new vacuum and the same reheating
as that after inflation will occur, see Fig.1. Thus it may be
expected that this thermalized region will be followed by a
standard evolution of radiation domination. In this note we will
check whether some of them are able to become our observable
universe.

These thermalized regions are generally different from the bubbles
nucleated by the CDL instanton. The bubbles are either empty or
dominated by the vacuum energy of new vacua, while the thermalized
regions are those filled with radiation and matter, which thus
more like some islands emerging from the dS background sea. Note
that in this note what is referred as the ``island'' is such an
emergently thermalized region which is generated by a large
fluctuation in original vacuum but emerges in a different vacuum,
see the green dashed line in Fig.1, which is actually slightly not
in its original meaning \cite{DV}. Thus here the emergence of
island will be inevitably related to the tunneling in the
landscape, especially the HM tunneling, as has been mentioned. In
this sense, the emerging probability of islands in new vacua may
be given by that of the HM instanton between different vacua,
while in Refs. \cite{DV, Piao0506} with the cosmological constant
background, it is not clear in what case we can calculate the
emerging probability of island.

The ``thermalized" here means that the resulting state is a
thermal state with radiation and matter, in which all components
are assumed to be in thermal equilibrium. Thus this part of region
is similar to that after the reheating following a slow roll
inflation. This can be distinguished from the discussion in Ref.
\cite{DKS}, in which the state after the fluctuation is assumed to
be an observable universe with structures, and also from the
recycling universe proposed in Ref. \cite{GV}, see also Ref.
\cite{LW1987}, in which the state after the fluctuation is another
dS spacetime.

\begin{figure}[t]
\begin{center}
\includegraphics[width=8cm]{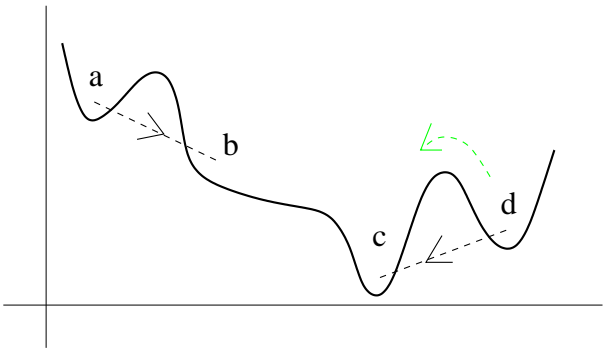}
\caption{The landscape of simple potential used to illustrate the
islands proposed here. In the eternal inflation, the bubbles with
vacuum `c' can be produced by the CDL instanton from either `a' or
`d' vacua. The endpoint of tunnelling from `a' would lie on a
plain region `b' of potential, and thus the bubble of
`a'$\rightarrow$`c' will have a period of inflation and then may
evolve to our real world, whereas the bubble of
`d'$\rightarrow$`c' will be empty, as is given by the black dashed
lines. However, here we pointed out that `d'$\rightarrow$`c' may
be also induced by some large and local quantum fluctuation with
the NEC violation, which produces some thermalized islands, as is
given by the green dashed line. These islands will evolve with the
standard FRW cosmology and some of them may like our observable
universes. Note that it is also possible that some
islands are generated in `b' due to the NEC violating fluctuations
in `a'. However, in this case the radiation and matter will be
diluted quickly and the island universe will then enter an
inflation phase dominated by the vacuum energy in `b', which will
have the same results as usual slow roll inflation. Here what we
concern is the case of the green dashed line, which might provides
a different avenue to an observable universe in the landscape. }
\end{center}
\end{figure}


Though the spawning of island in the landscape is actually a
quantum process, it may be regarded phenomenally or
semiclassically as a NEC violating evolution to study, as was done
in Ref. \cite{Piao0506}, see also Ref. \cite{PZ}. Thus in this
sense the island actually shares some remarkable successes of
inflation model. The reason is that the inflation can be generally
regarded as an accelerated stage, and so may defined as an epoch
when the comoving Hubble length decreases, which actually occurs
equally during a NEC violating expansion. When the island emerges,
the change of local background may be depicted by the drastic
evolution of local Hubble parameter `$h$', where the ``local"
means that the quantities, such as the scale factor `$a$' and
`$h$', only character the values of the NEC violating region. We
begin with introducing `$\epsilon$' defined $-\,{\dot h}/ h^2$,
which can be regarded as $\epsilon \simeq {1\over h \Delta
t}{\Delta h\over h}$, and thus actually describes the change of
$h$ in unit of Hubble time. During the NEC violating fluctuation,
${\dot h}>0$, thus $\epsilon <0$ can be deduced. We assume here
that $\epsilon $ is constant for simplicity. Thus after making the
integral for the definition of $\epsilon$, we have $a\sim h^{1/
|\epsilon |}$.



The more rapid the fluctuation is, in principle the stronger it
can be, which in some sense is also a reflection of the
uncertainty relation between the energy and time in quantum
dynamics. Therefore to make the NEC violating fluctuate be so
strong as to be able to create the islands of our observable
universe, we should take the time scale of the NEC violation \be
\int dt = \int_i^e {dh \over |\epsilon | h^2} \simeq {1\over
|\epsilon | h_i} \label{indt}\ee be vanishingly small, where the
subscript `i' and `e' denote the initial and end value of the NEC
violating fluctuation respectively, and thus $h_i$ is determined
by the energy scale of original vacuum. To make $\int dt
\rightarrow 0$, we need $h_i$ or $|\epsilon |$ is quite large. The
larger $h_i$ is the more violent the fluctuation is anticipative.
However, as will be showed here, $h_i$ is required to be so small
as to solve the horizon problem of standard cosmology. Thus
Eq.(\ref{indt}) suggests $|\epsilon |\gg 1$, which means that
though during the fluctuation the change of $h$ is drastic, the
expansion of the scale factor is extremely slow, since $a\sim
h^{1/|\epsilon|}$.





The scale of the NEC violating region is generally required to be
larger than the Hubble scale of original vacuum \cite{DV}, see
also Refs. \cite{FG, VT, BTV}. This is also consistent with
application of the HM instanton action in which the region
tunnelling to the top of the barrier corresponds to the Hubble
scale of the original vacuum, which may be understood by using the
stochastic approach to inflation \cite{AAS, GLM, L90, L92}. This
result sets the initial value of local evolution of $a$, and since
it is nearly unchanged during the fluctuation, \be a_e\simeq
a_i\simeq 1/h_i, \label{ae}\ee may be deduced, which means that
the smaller $h_i$ is, the larger the scale of local thermalized
region after the fluctuation is. To obtain enough efolding number
for solving the horizon problem of standard cosmology, we need
$a_e\gg 1/h_e$, thus $h_i\ll h_e$ is required, which may be also
seen as follows. \be {\cal N} \equiv \ln{\left({a_e h_e\over a
h}\right)} \label{caln}\ee is the efolding number of mode with
some scale $\sim 1/k$, where $k=ah$, which leaves the horizon
before the end of the NEC violating fluctuation, and thus $k_e$ is
the last mode to be generated. When taking $ah=a_0h_0$, where the
subscript `0' denotes the present time, we generally have ${\cal
N}\sim 50$, which is required by observable cosmology, see Ref.
\cite{LL} for a discussion on the value of $\cal N$. By using
Eq.(\ref{ae}) and (\ref{caln}), we may obtain approximately \be
{\cal N}\simeq \ln\left({h_e\over h_i}\right)\simeq
\ln\left({T_e\over \Lambda_i}\right)^2, \label{caln1}\ee where
$\Lambda_i\simeq h_i^{1/2}$ is the energy scale of original vacuum
and ${T}_e\simeq h_e^{1/2}$ is the thermalized temperature after
the NEC violating fluctuation, which may be the same as the
reheating temperature after inflation, and $m_p^2=1$ has been
taken. When taking ${\cal N}\simeq 50$ and ${T}_e \sim 10^{15}$
Gev, we have $\Lambda_i\sim$ Tev. For a lower ${T}_e$, $\Lambda_i$
is required to be smaller. Thus unlike the case in Refs. \cite{DV,
Piao0506}, it seems that here the efolding number required to
solve the horizon problem of standard cosmology can not be always
obtained. To have an enough efolding number, an enough low
original vacuum should be selected. The reason is that the smaller
$\Lambda_i$ is, the larger the Hubble scale of corresponding dS
vacuum is, and thus the size of local universe after the
fluctuation and the efolding number. This result also suggests
that if our universe is actually such an island originated from
last vacuum, then the observations made in our universe might have
recorded some information on last vacuum. For example, for the
efolding number ${\cal N}>50$, we may know that its energy scale
should be low, and further if the thermalization temperature $T_e$
is about $10^{15}$ Gev, our universe must not be originate from a
fluctuation of those vacua with the energy density larger than Tev
scale. In principle, the $T_e$ should be required to be lower than
that the monopoles production needs, while higher than Tev. This
can not only avoid the monopole problem afflicting the standard
cosmology, but helps to provide a solution to the matter genesis.

The calculations of primordial scalar perturbation during the
evolution with the NEC violation have been done in Refs.
\cite{Piao0506, PZ, Piao0705, Piao0512}. The emergence of island
in the landscape corresponds to the limit case with $\epsilon \ll
-1$.
In Ref. \cite{Piao0506}, which firstly calculates the curvature
perturbation of island universe, in which the background vacuum is
taken as the observed value of cosmological constant, it has been
shown that the spectrum of Bardeen potential $\Phi$ before the
thermalization is dominated by an increasing mode and is nearly
scale invariant, which under some conditions may induce scale
invariant curvature perturbation. Whether the resulting spectrum
is scale invariant is determined by the physics at the epoch of
thermalization. Thus in this case there is generally an
uncertainty, see the discussions in Ref. \cite{Piao0506}. However,
later it was noted that the curvature perturbation may be also
induced by the entropy perturbation \cite{Piao0705}, or by the
perturbation of test scalar field \cite{Piao0512} before the
thermalization, which under certain conditions may has a nearly
scale invariant spectrum and proper amplitude required by the
observations. The curvature perturbation induced by the entropy
perturbation has same form as that by the Bardeen potential
\cite{Piao0705}, only up to a numerical factor with unite order,
which, more importantly, is not dependent of the physical detail
of thermalized surface. Thus unlike the case in Refs. \cite{DV,
Piao0506}, when with more freedom degrees provided by the
landscape, since in low energy the landscape may be approximated
as the space of a set of fields with a complicated and rugged
potential, it may be more natural to consider the curvature
perturbation induced by the entropy perturbation, which will
definitely give the scale invariant spectrum with proper
amplitude. In Refs. \cite{Piao0506, Piao0705}, the amplitude of
curvature perturbation is given by \be {\cal P}_{s} \cong
|\epsilon |h_e^2, \label{pxi}\ee which only depends on $|\epsilon
|$ and $h_e$. When $|\epsilon |\rightarrow \infty$, the
perturbation amplitude will be divergent. Thus for our purpose,
the value of $|\epsilon |$ seems to require a litter fine tuning.
For example, having taken $T_e\sim 10^{15}$ Gev as the
thermalization temperature and $|\epsilon |\sim 10^2$, we have
${\cal P}_s \sim 10^{-10}$ for (\ref{pxi}), which is just the
observed amplitude of CMB \cite{WMAP}. In principle, $\epsilon$
may be taken as a larger value, however, in this case to have a
proper amplitude, the thermalization temperature $T_e$ should be
smaller. Here the fine tuning for $\epsilon$ is actually similar
to that appearing in the inflation model. They are related to each
other by a dual transformation $|\epsilon|\leftrightarrow
1/|\epsilon|$ \cite{Piao0705}, which also corresponds to a duality
between their background evolutions, i.e. between the nearly
exponent expansion with $|\epsilon|\simeq 0$ and the slow
expansion $|\epsilon |\rightarrow \infty$.

The calculations of primordial tensor perturbation during the
evolution with the NEC violation have been done in Ref.
\cite{Piao0601}. The case corresponding to the island universe is
given in Eq.(13) of Ref. \cite{Piao0506}. The spectrum is \be
{\cal P}_{\rm T} \cong k^3\left|{v_k^{(e)}\over a_e}\right|^2
\simeq {k^2\over a_e^2}, \label{ptt}\ee where the gauge invariant
variable $v_k$ is related to the tensor perturbation $h_k$ by $v_k
= a h_k$. It can be seen that the tensor spectrum is quite blue,
which is actually a reflection of the rapid increase of background
energy density during the NEC violating fluctuation, see Ref.
\cite{Piao0601} for details. This result indicates that the tensor
amplitude in the island will be intensely suppressed on large
scale. To calculate the tensor amplitude, we may replace $a_e$ by
using $k_e=a_eh_e$ in Eq.(\ref{ptt}) and have \be {\cal P}_{\rm T}
\simeq h_e^2\left({k\over k_e}\right)^2, \label{pt}\ee which gives
the value of tensor amplitude in various scales. Thus the value on
large scale may be obtained by combining Eqs.(\ref{caln}) and
(\ref{caln1}), and then substituting the result into
Eq.(\ref{pt}), which is ${\cal P}_{\rm T} \cong h_i^2$. This is
quite low, for example, taking $h_i^2\simeq \Lambda_i\sim $Tev, we
have ${\cal P}_{\rm T} \sim 10^{-60}$. Note that this result on
large scale is same as that in Ref. \cite{DV}, in which the dS sea
phase is suddenly matched to a radiation dominated phase and the
NEC violating mediated phase is neglected. However, in their paper
they argued that the tensor spectrum is scale invariant and thus
has same amplitude in all scales, while in our calculation the
spectrum is actually strong blue, which only is same as that in
Ref. \cite{DV} on large scale. The discrepancy between Refs.
\cite{DV} and \cite{Piao0506} lies in the NEC violating phase
neglected by Ref. \cite{DV}, which it is that tilts the tensor
spectrum. In small scale, which corresponds to take $k=k_e$, we
have ${\cal P}_{\rm T} \simeq h_e^2$, which is actually quite
large for a high thermalization scale.


In summary, we point out that in the eternal inflation driven by
the metastable vacua of the landscape, it may be possible that
some large and local quantum fluctuations with the NEC violation
can stride over the barriers between different vacua in the
landscape and straightly create some thermalized islands in new
vacua. We show that these islands may be consistent with our
observable universe. This result suggests that with the landscape
the observable universe may be some of many thermalized regions,
which appear either by a slow roll inflation after the nucleation
of bubbles, followed by the reheating, or by a straightly
thermalization in new vacua without the slow roll inflation,
spawned within the eternally inflating background.

The observations in principle can determined whether we live in an
emergently thermalized island or in a reheating region after
inflation inside bubble. It has been shown that in the island the
tensor amplitude is negligible on large scale, while there exists
a large class inflation model, such as large field inflation
model, with moderate amplitude of tensor perturbation, see e.g.
Ref. \cite{LR} for the various inflation models. Thus it seems
that the detection of a stochastic tensor perturbation will be
consistent with the inflation model,
while rule out the possibility that an straightly thermalized
region is regarded as our real world.
The low tensor amplitude on large scale is also not conflicted
with the inflation model, e.g. some small field inflation models.
Thus in this case other distinguishabilities need to be
considered. The bubble after the nucleation described by the CDL
instanton is generally negatively curved, and thus the
corresponding universe is an open universe, while the island
leaded to by the NEC violating fluctuation may be closed.
Thus in principle if the cosmological dynamics is actually
controlled by a landscape with many metastable vacua, the
curvature measurement of our universe will be significant to make
clear where we live in.

It should be fairly said that the discussions here is inevitably
slightly speculative, since a full description for the phenomena
with the NEC violation is still lacked for the moment. However,
the results showed here might have captured some essentials of
emergently thermalized island in the landscape, which might be
interesting and significant to phenomenological study of landscape
cosmology.

\textbf{Acknowledgments} This work is supported in part by NNSFC
under Grant No: 10405029, in part by the Scientific Research Fund
of GUCAS(NO.055101BM03), in part by CAS under Grant No:
KJCX3-SYW-N2.

\end{document}